\documentclass[]{spie}  

\usepackage{amsmath,amsfonts,amssymb}
\usepackage{graphicx}
\usepackage[colorlinks=true]{hyperref}
\usepackage[]{threeparttable}

\def\arcsec{\hbox{$^{\prime\prime}$}}

\def\farcm{\hbox{$.\mkern-4mu^\prime$}}

\def\gtrsim{\mathrel{\hbox{\rlap{\hbox{%
  \lower4pt\hbox{$\sim$}}}\hbox{$>$}}}}

\date{July 19, 2016 draft for SPIE Astronomical Telescopes + Instrumentation 2016}
\title{Slitless spectroscopy with the James Webb Space Telescope Near-Infrared Camera (JWST NIRCam)}

\author[a]{Thomas P. Greene}
\author[b]{Laurie Chu}
\author[c]{Eiichi Egami}
\author[b]{Klaus W. Hodapp}
\author[c]{Douglas M. Kelly}
\author[c]{Jarron Leisenring}
\author[c]{Marcia Rieke}
\author[d]{Massimo Robberto}
\author[c]{Everett Schlawin}
\author[d]{John Stansberry}

\affil[a]{NASA Ames Research Center, Moffett Field, CA, USA}
\affil[b]{Institute for Astronomy, University of Hawai'i, Honolulu, HI, USA}
\affil[c]{Steward Observatory, University of Arizona, Tucson, AZ, USA}
\affil[d]{Space Telescope Science Institute, Baltimore, MD, USA}

\authorinfo{Send correspondence to T.P.G.: E-mail: tom.greene@nasa.gov, Telephone: +001 650 539 5244  
}
\begin{document} 
\maketitle

\begin{abstract}
The James Webb Space Telescope near-infrared camera (JWST NIRCam) has two $2\farcm2 \times 2\farcm2$ fields of view that are capable of either imaging or spectroscopic observations. Either of two $R \sim 1500$ grisms with orthogonal dispersion directions can be used for slitless spectroscopy over $\lambda = 2.4 - 5.0$ $\mu$m in each module, and shorter wavelength observations of the same fields can be obtained simultaneously. We present the latest predicted grism sensitivities, saturation limits, resolving power, and wavelength coverage values based on component measurements, instrument tests, and end-to-end modeling.  Short wavelength (0.6 -- 2.3 $\mu$m) imaging observations of the 2.4 -- 5.0 $\mu$m spectroscopic field can be performed in one of several different filter bands, either in-focus or defocused via weak lenses internal to NIRCam. Alternatively, the possibility of 1.0 -- 2.0 $\mu$m spectroscopy (simultaneously with 2.4 -- 5.0 $\mu$m) using dispersed Hartmann sensors (DHSs) is being explored. The grisms, weak lenses, and DHS elements were included in NIRCam primarily for wavefront sensing purposes, but all have significant science applications. Operational considerations including subarray sizes, and data volume limits are also discussed. Finally, we describe spectral simulation tools and illustrate potential scientific uses of the grisms by presenting simulated observations of deep extragalactic fields, galactic dark clouds, and transiting exoplanets.
\end{abstract}

\keywords{James Webb Space Telescope, JWST, NIRCam, grisms, slitless spectroscopy}

\section{INTRODUCTION}
\label{sec:intro}  

The near-infrared camera (NIRCam) of the James Webb Space Telescope
(JWST) has numerous flexible modes that provide wide-field imaging
through narrow, medium or wide filters, coronagraphic imaging, and
slitless spectroscopy\cite{BEG12, KBT07, RKH05, GBE07}. These and other
previous works have described its wide-field and coronagraphic imaging
capabilities well, but little practical information has been published
on using NIRCam in its spectroscopic modes. Long-wave (LW; $\lambda =
2.4 - 5.0$ $\mu$m) Si grisms\cite{JWM08} and short-wave (SW; $\lambda =
1 - 2$ $\mu$m) Dispersed Hartmann Sensors (DHSs) were developed for the
purposes of wavefront sensing and telescope primary segment phasing
with NIRCam\cite{SKS08}. The LW grisms have been approved and are being
supported for scientific use. The SW DHSs are not yet supported but are
being considered for scientific use during Cycle 1 and later
observations.

Both of these optics give NIRCam the ability to perform slitless
spectroscopic observations. Full-field slitless observations produce
spectral images that contain a spectrum for nearly all objects in the
NIRCam imaging field. This is more complete spatially that using a
spectrometer with apertures (e.g., opening the MSA masks in the JWST
NIRSpec instrument) and is more suitable for mapping regions of the sky
(e.g., to search for high redshift emission line objects). When
implemented in a camera like NIRCam, the spectral bandpass of slitless
exposures can be selected with filters. This can be used to limit the
content (e.g., spectral features or redshift range) as well as optimize
the sensitivity and extent of spectra. Overlapping spectra of multiple
objects can often be resolved by taking separate slitless spectral
exposures with orthogonal dispersion directions. Slitless spectroscopy
is also valuable for high precision observations of bright stars (e.g.,
time series spectra of transiting planets) because there is no slit to
impart artifacts into the spectrum in the presence of telescope jitter
or if there are variations in the PSF with wavelength (e.g.,
diffraction-limited images). Slitless spectra do see the full
background of their undispersed bandpasses and therefore have lower
sensitivities than those of slit spectrographs. However, the low
background of JWST allows NIRCam slitless spectra to have sensitivities
within a factor of a few of NIRSpec in the $\lambda = 2.4 - 5.0$ $\mu$m
range.

The NIRCam instrument is composed of two nearly identical optical
modules (A and B), each having a $2\farcm2 \times 2\farcm2$ field of
view (FOV). Wavelengths $\lambda < 2.4$ $\mu$m are imaged by the
short-wavelength (SW) channel of each module which has a focal plane of
four butted 2048 $\times$ 2048 (2040 $\times$ 2040 active) pixel
HAWAII-2RG detectors with a plate scale of 32 mas pixel$^{-1}$. A
dichroic beamsplitter sends longer wavelengths to the long-wavelength
(LW) channel that has a single 2048 $\times$ 2048 (2040 $\times$ 2040
active) pixel HAWAII-2RG detector focal plane with a plate scale of 65
mas pixel$^{-1}$ \cite{BEG12, RKH05}. This allow simultaneous SW and LW
imaging of the same NIRCam field with each module. The pupil wheels in
the LW channels of each module are equipped with two identical Si
grisms with perpendicular orientations so that light is dispersed along
either the rows (R grisms) of columns (C grisms) of their detectors.
Each module also has two SW pupil wheel positions containing a DHS with
10 apertures, with each aperture containing a grism and a wedge to
spatially separate the 10 spectra. The dispersions of all 10 DHS grisms
in each wheel position are oriented either along rows (DHS0) or else
rotated 60$^\circ$ (DHS60)\cite{SKS08}. Each grism in each DHS is
mapped to a pair of telescope segment edges and in aggregate subtend
approximately 40\% of the geometric telescope pupil area.

We present practical information on using NIRCam's spectroscopic modes
in the following sections. First we present basic spectroscopic
capabilities and performance limits of the grisms and DHS elements.
Next we discuss operational considerations including using these
elements simultaneously. We conclude with simulations of wide-field
multi-object spectra of deep extragalactic fields and galactic dark
clouds as well as single-object simulations of transiting exoplanet
spectra.

\section{Spectroscopic Capabilities} \label{capabilities}

\subsection{Basic Performance Parameters}

The LW grisms were designed to have undeviated wavelengths of 4.0 $\mu$m and to provide a dispersion of 10.0 \AA~ pixel$^{-1}$ in NIRCam\cite{GBE07}. Dispersions for all four grisms (two in each module) were measured to be within 1\% of this value during JWST science instrument testing at NASA Goddard Space Flight Center (GSFC) during the summer of 2014.

\begin{figure*}
\begin{center}
\includegraphics[width=0.28\textwidth]{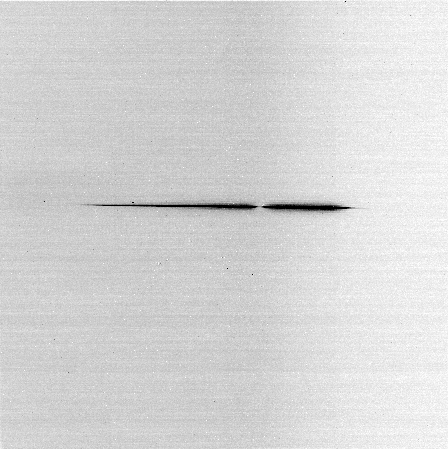}
\includegraphics[width=0.37\textwidth]{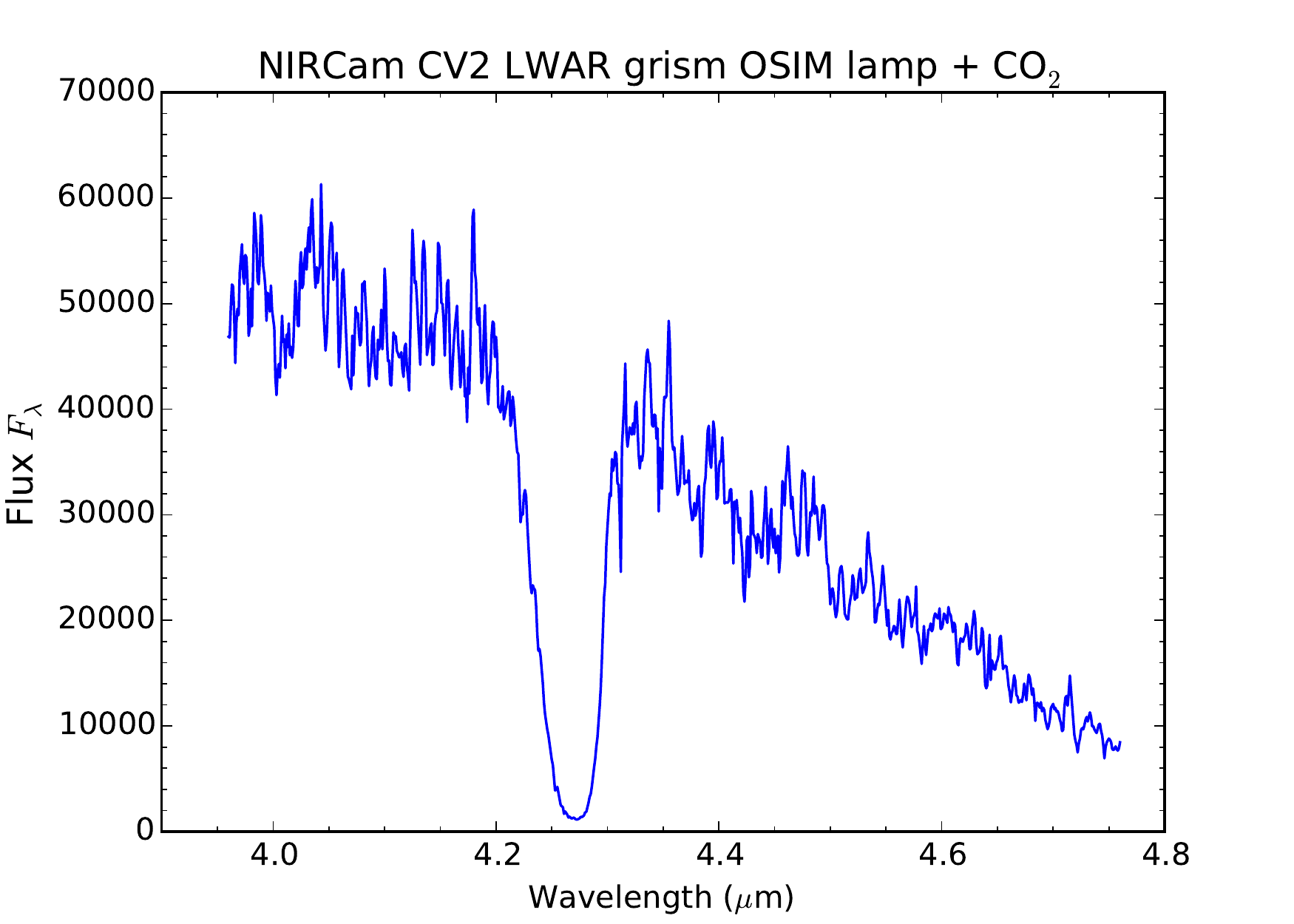}
\includegraphics[width=0.33\textwidth]{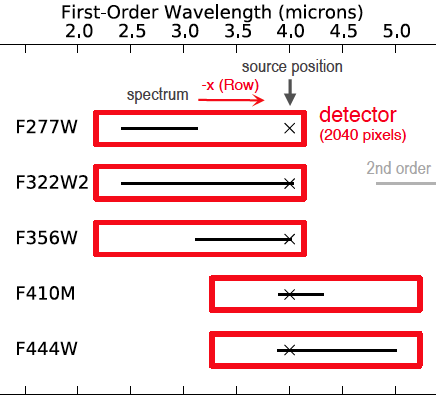}
\end{center}   
     \caption{ \label{fig:grism} 
\small  Left: NIRCam spectral image of the OSIM super-continuum lamp taken with the LWA R grism and F444W filter during JWST instrument instrument testing at NASA GSFC in August 2014. Wavelength increases to the left (-X direction). Center: Extracted spectrum from the image with an approximate wavelength calibration applied. The continuum decreases toward longer wavelengths due to low fiber transmittance, and the broad feature near 4.27 $\mu$m is due to CO$_2$ absorption. Both of these features are artifacts of the test equipment and not NIRCam itself. Right: Layout of an object's LW R grism spectra spectrum relative to its direct image position for different filters (image courtesy of D. Coe). }
\end{figure*}

Filters are used in series with grisms to select the desired spectral range of the observation. The NIRCam LW filter wheel contains filters that are medium ($\lambda / \Delta \lambda \sim 10 - 20$), wide ($\lambda / \Delta \lambda \sim 4$), and double-wide ($\lambda / \Delta \lambda \sim 2$) bandpasses\footnote{see http://www.stsci.edu/jwst/instruments/nircam/instrumentdesign/filters/}, and any of these can be used with the grisms in principle. A subset of these is being defined for science applications. All LW narrow-band filters reside in the same pupil wheel as the grisms, so these cannot be used in series with the grisms. Each grism spans the entire telescope pupil, and any source in the NIRCam imaging field will have its light dispersed by a grism that is selected in the pupil wheel. Some of the dispersed spectrum may land outside of the FOV and not be recorded by the detector depending on the object field position and the selected filter (See Section \ref{Operations}). The image and extracted spectrum of a NIRCam Module A LW grism exposure taken during the JWST instrument test campaign at NASA GSFC in 2014 is shown in Figure~\ref{fig:grism}. This figure also shows the layout of an object's LW R grism spectra spectrum relative to its direct image position for different filters.

\begin{figure*}
\begin{center}
\includegraphics[width=0.33\textwidth, angle=0]{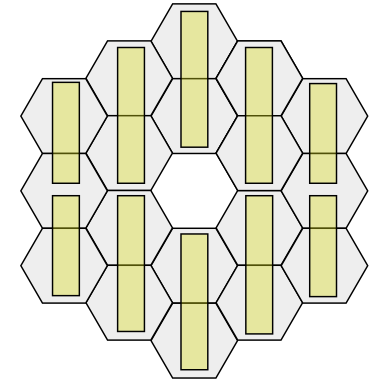}
\includegraphics[width=0.65\textwidth]
     {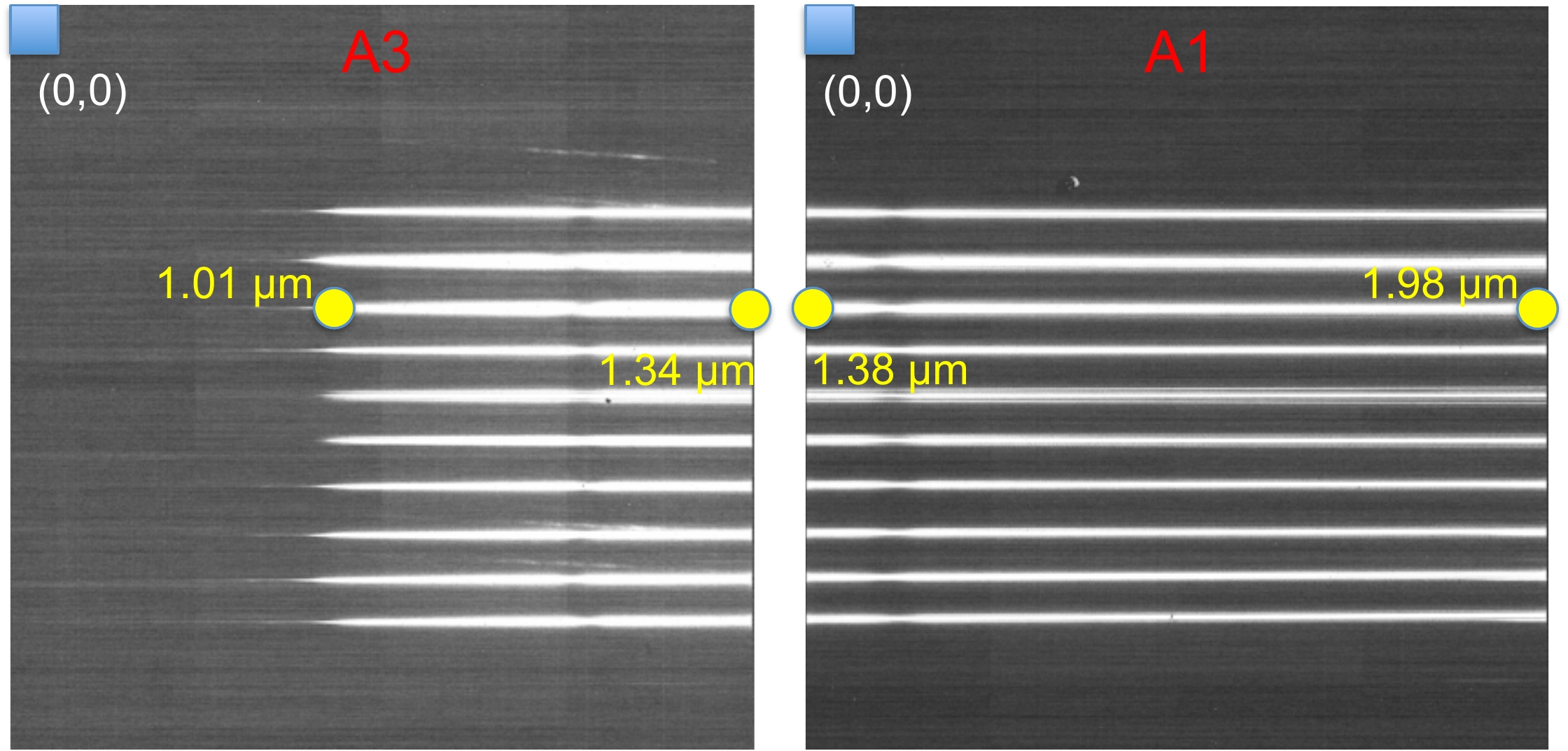}
\end{center}   
     \caption{ \label{fig:DHS} 
\small  Left: Schematic projection of the SW DHS0 elements onto the JWST primary mirror segments. Each of the 10 DHS elements span 2 segment edges, and approximately 40\% of the total optical telescope element (OTE) aperture is seen by all DHS elements in total. Obscurations by the secondary supports are not shown. A NIRCam image of all 10 DHS0 $\lambda = 1 - 2$ $\mu$m spectra is also shown, with the shorter wavelength A3 detector image in the center panel and the longer wavelength A1 detector image in the right panel. Wavelength increases from left to right in both detector images. The source was positioned such that the 1.01 $\mu$m short half-power wavelength of the series F150W2 filter appears near the center of the A3 detector (center panel), and its 2.33 $\mu$m long half-power wavelength falls beyond the right edge of the A1 detector (right panel). The gap between the 2 detectors is approximately 150 pixels ($\sim 5\arcsec$ on the sky), corresponding to about 0.045 $\mu$m (45 nm) in wavelength.
}
\end{figure*}

The SW DHS grisms have undeviated wavelengths of 1.36 $\mu$m, and
their dispersions have been measured to be 2.90 \AA (Module A) and 2.91
\AA (Module B) per NIRCam SW pixel. Therefore two NIRCam SW filters
plus their $\sim 5\arcsec$ gap span approximately 1.235 $\mu$m in
wavelength. DHS elements will be typically used with F150W2 filters
which have bandpasses from 1.01 to 2.33 $\mu$m, so $\lambda = 1 - 2$
$\mu$m DHS spectra can be imaged simultaneously on 2 adjacent SW
detectors. Observing an object with the DHS requires that it be located
in a specific region of the NIRCam field to ensure that all 10 DHS
spectra are imaged onto the SW detectors over their entire $\lambda = 1
- 2$ $\mu$m regions. Multiple objects in the field can also produce
overlapping DHS spectra. Therefore DHS observations are best suited for
single-object spectroscopy, and relatively bright stars are required to
achieve high signal-to-noise. Figure~\ref{fig:DHS} shows the layout of
the DHS0 elements on the JWST pupil and an image of the resultant 10
DHS spectra.

Both the LW grisms and DHSs are usually operated in first order (m = 1) where they have maximum efficiency. Their high dispersions, the finite size of the NIRCam field and detectors, and the limited bandpasses of the series filters all limit the detection of other orders. It is only possible to detect m $\neq$ 1 light when using the F322W2 filter with an LW grism. In that case, an object at the right location can produce a spectrum that is m = 1 at long wavelengths and m = 2 at short wavelengths (relative wavelengths within the F322W2 bandpass; see Figure~\ref{fig:grism} right panel). The broad bandpass of the F150W2 filter will cause DHS $\lambda < 2$ $\mu$m, m = 2 spectra to overlap partially with $\lambda > 2$ $\mu$m, m = 1 DHS spectra. In particular, m = 1, 2.02 -- 2.33 $\mu$m wavelengths will be contaminated with m = 2 light from 1.01 -- 1.165 $\mu$m wavelengths.

\subsection{Performance: Resolution, Sensitivities and Saturation Limits}

\begin{figure*}
\begin{center}
\includegraphics[width=0.49\textwidth, angle=0]{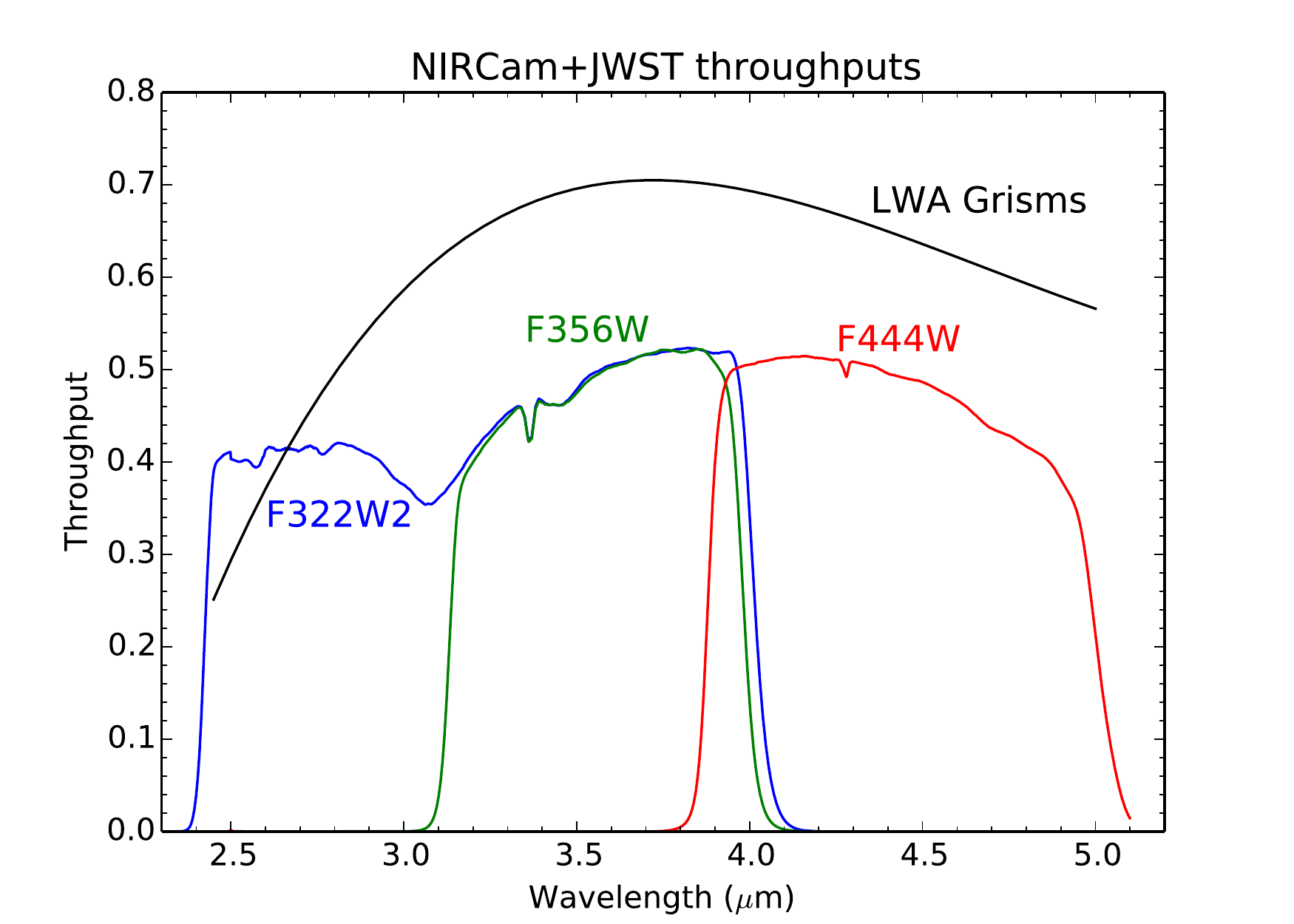}
\includegraphics[width=0.49\textwidth]{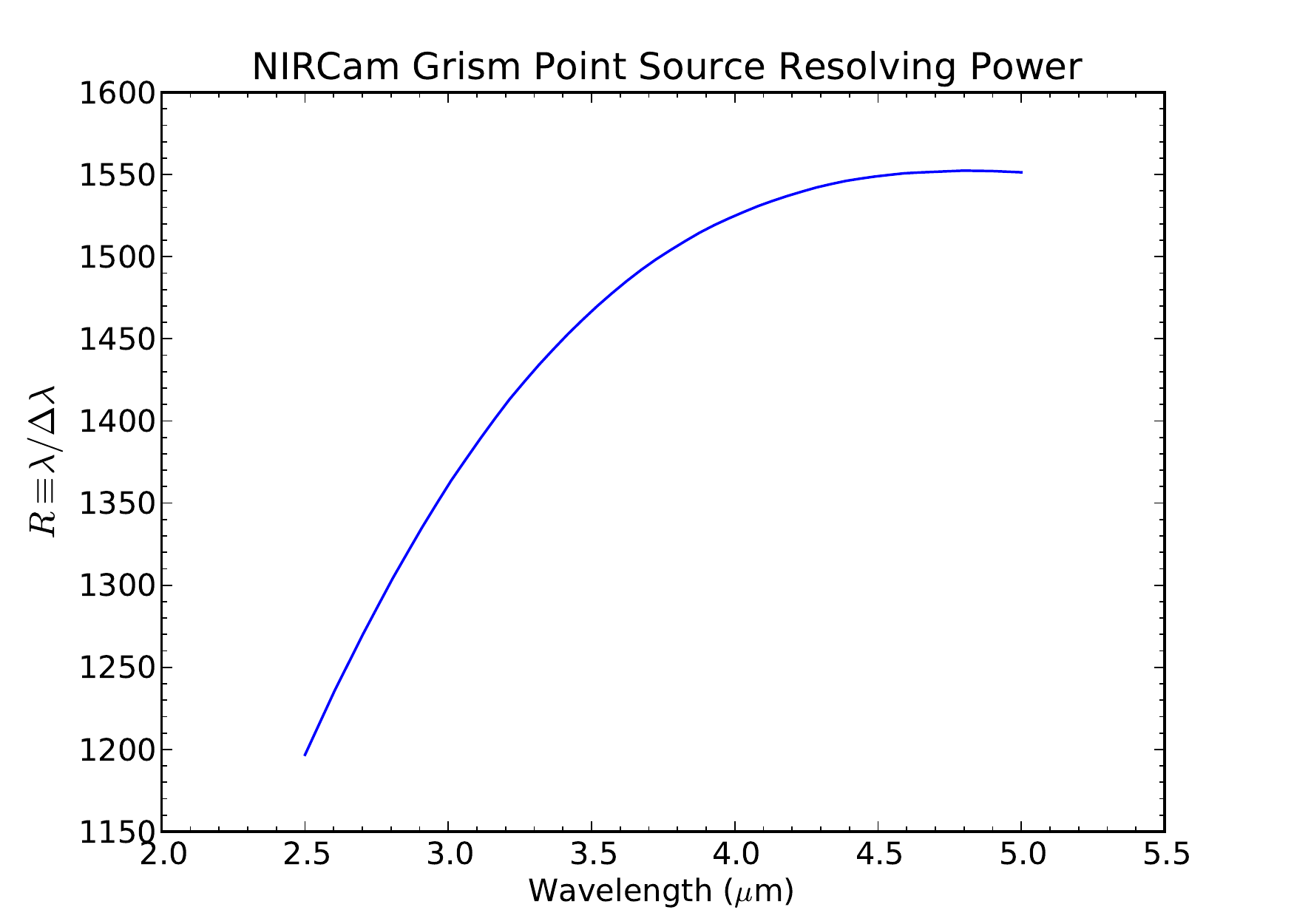}
\end{center}   
     \caption{ \label{fig:tput_and_R} 
\small  Left: Total system throughput including all OTE and NIRCam optics and the detector quantum efficiency for several NIRCam filters. The theoretical LW grism efficiency curve (shown for the A module) must be multiplied by the filter curves to produce the system throughput at each wavelength. The Module B LW grisms are anti-reflection coated on only 1 side and therefore have throughputs approximately 25\% lower than the LWA grisms. Right: Grism FWHM spectral resolving power vs. wavelength for point sources, limited by pixel sampling of the PSF at shorter wavelengths ($\lambda \lesssim 4$ $\mu$m) and limited by the circular beam factor\cite{ER00} and diffraction at longer wavelengths ($\lambda \gtrsim 4$ $\mu$m).}
\end{figure*}

The transmission and reflectance values of all NIRCam optical components used for imaging observations have been tested, and the composite throughput values for Module A (including the OTE reflectivity as well as NIRCam detector quantum efficiency) are shown in Figure~\ref{fig:tput_and_R}. This figure also shows the theoretical first-order Module A LW grism efficiency, which was validated with measurements at two wavelengths near 3 $\mu$m during pre-delivery instrument testing at the Lockheed Martin Advanced Technology Center. The grism efficiency must be multiplied by the NIRCam + OTE curve for the chosen series filter to produce the system throughput at each wavelength. Module A LW grisms are anti-reflection coated on both sides, while those in the B module are coated only on their flat (non-grooved) sides. Therefore the Module B LW grism efficiencies are $\sim 0.74$ times that of the Module A ones, and the Module B ones produce some ghosts of bright spectra.

All NIRCam spectroscopy is conducted without using slits or other focal plane apertures, so the spectral resolving power $R$ of its grism or DHS observations are determined by the dispersion, wavelength, 
its circular pupil\cite{ER00}, detector sampling, and object size. The basic equation for point-source resolving power $R \equiv \lambda / \Delta \lambda$ dictates that $R$ increases with wavelength for a constant spectral bandpass $\Delta \lambda$ (e.g., an undersampled 2 pixel PSF), but this is flattened by diffraction for the LW grisms at wavelengths $\lambda \gtrsim 4$ $\mu$m where the well-sampled NIRCam PSF causes point sources to subtend more pixels. The resultant spectroscopic resolving power for LW grism observations of point sources is shown in the right panel of Figure~\ref{fig:tput_and_R}. The pixel-limited resolving power (strongest at $\lambda \lesssim 4$ $\mu$m) may be improved with dithering of multiple observations.

Table~\ref{tab:grismA} gives the 10 $\sigma$ point-source continuum
sensitivities for the Module A grisms using a 2 (spectral) $\times$ 5
(spatial) pixel extraction aperture in 10,000 s of integration time.
Unresolved point source emission line sensitivities are also given,
using the spectral resolving power computed for each wavelength (right
panel of Figure~\ref{fig:tput_and_R}). All sensitivities were
calculated for medium zodiacal background levels using either the
F322W2 or F444W filter in series, as identified in the table. These
sensitivities are a factor of $\sim5$ worse than expected for the JWST
NIRSpec instrument\footnote{see
http://www.stsci.edu/jwst/instruments/nirspec/sensitivity/}, but
NIRSpec cannot be operated in a true slitless mode. Lower background
conditions than those assumed here will give sensitivities that are up
to a factor of $\sim 1.4$ lower (better), and higher background
conditions will result in sensitivities that are up to a factor of
$\sim 1.4$ higher (worse). Using narrower band filters (i.e., F356W
instead of F322W2, or F430M instead of F444W) can improve sensitivities
by a factor of $\sim 1.2 - 2$. Table~\ref{tab:grismA} also gives
saturation values for 0.68 s integrations, 2 frame times for 2048
$\times$ 64 pixel subarrays with the detector operated with four
outputs (stripe mode). Larger subarray stripes (2048 $\times$ $>$64
pixels) will result in fainter bright limits, and operating in window
mode (using a single detector output) will make the bright limits 1.5
mag fainter also. The Module B grisms will have sensitivities
approximately 1.16 times higher (worse) and saturation limits 0.33 mag
brighter than the ones tabulated for Module A.

\begin{table}
\centering
\begin{threeparttable}[ht]
\caption{Module A Grism performance table\tnote{a}} 
\label{tab:grismA}    
\begin{tabular}{|r|r|r|r|r|l|} 
\hline
\rule[-1ex]{0pt}{3.5ex}  $\lambda$ ($\mu$m) & F$_{\rm cont}$ ($\mu$ Jy)\tnote{b} & F$_{\rm line}$ (W m$^{-2}$)\tnote{c} & $K_{\rm sat}$ (A0V)\tnote{d} & $K_{\rm sat}$ (M2V)\tnote{d} & Filter\tnote{e}\\
\hline
2.5  & 11.1 & 1.09E-20 & 4.3 & 4.3 & F322W2 \\ \hline
2.7  & 8.7	& 7.35E-21 & 4.5 & 4.6 & F322W2 \\ \hline
2.9  & 8.0	& 5.98E-21 & 4.3 & 4.5 & F322W2 \\ \hline
3.1  & 7.9	& 5.22E-21 & 4.2 & 4.4 & F322W2 \\ \hline
3.3  & 6.7	& 3.97E-21 & 4.2 & 4.5 & F322W2 \\ \hline
3.5  & 6.5	& 3.45E-21 & 4.0 & 4.3 & F322W2 \\ \hline
3.7  & 6.3	& 3.05E-21 & 3.9 & 4.2 & F322W2 \\ \hline
3.9  & 7.0	& 3.11E-21 & 3.6 & 3.9 & F322W2 \\ \hline
4.1  & 12.1 & 4.99E-21 & 3.5 & 3.8 & F444W \\ \hline
4.3  & 13.5 & 5.18E-21 & 3.2 & 3.5 & F444W \\ \hline
4.5  & 15.1 & 5.38E-21 & 2.9 & 3.0 & F444W \\ \hline
4.7  & 19.1 & 6.38E-21 & 2.5 & 2.7 & F444W \\ \hline
4.9  & 25.1 & 7.88E-21 & 2.2 & 2.3 & F444W \\

\hline
\end{tabular}
\begin{tablenotes}
      \small
	  \item [a] Module B grisms will have sensitivities approximately 1.16 times higher (worse) and saturation limits 0.33 mag brighter.
	  \item [b] 10 $\sigma$ point-source continuum sensitivities for 10,000 s integrations using a 2 (spectral) $\times$ 5 (spatial) pixel extraction aperture. 
	  \item [c] 10 $\sigma$ point-source unresolved emission line sensitivities for 10,000 s integrations using the actual spectral resolving power at this wavelength (right panel of Figure~\ref{fig:tput_and_R}).
	  \item [d] K-band magnitudes for saturation (80\% full well or 65,000 electrons) for 0.68 s integrations (2 reads) of 2048 $\times$ 64 pixel regions in stripe mode (4 outputs). Larger subarrays or use of only 1 output will result in fainter bright limits.
      \item [e] Narrower filters will have similar saturation values and somewhat lower (better) sensitivities.
    \end{tablenotes}
\end{threeparttable} 
\end{table}

As seen in Figure~\ref{fig:DHS}, the DHS elements subtend about 40\% of the OTE pupil, and they have transmission values on the order of 0.6. This and the fact that each DHS component produces a separate spectrum results in relatively low photon fluxes in each detector pixel. Consequently, full-array DHS exposures with 2 frame integrations (21.4 s integration times) have a saturation limit (80\% full well) of 5.0 -- 3.25 mag (Vega) for a G2V star over the 1.0 -- 2.0 $\mu$m wavelength range, respectively. This improves to 1.2 -- -0.3 mag when only a 2048 $\times$ 64 pixel detector region is operated in stripe mode (four outputs), but this configuration will record only 1 or 2 of the 10 DHS spectra.\\

\section{Spectroscopic Operations} \label{Operations}

\begin{figure*}
\begin{center}
\includegraphics[width=0.9\textwidth,angle=0]{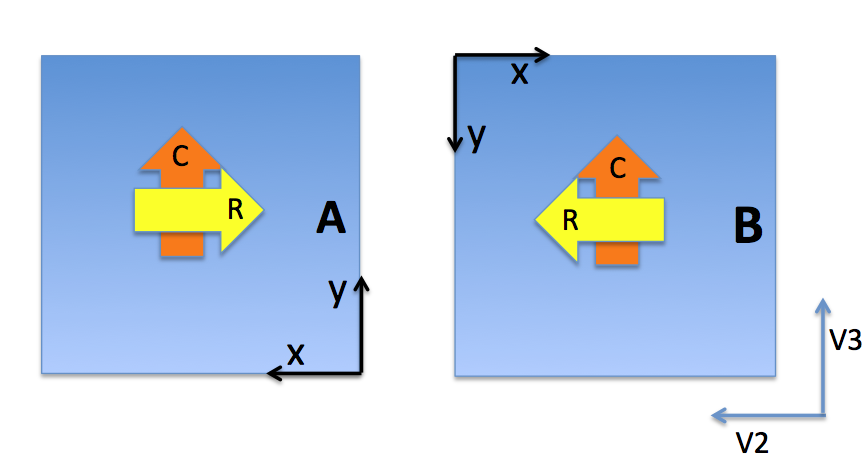}
\end{center}
     \caption{ \label{fig:FP_layout} 
\small  Grism dispersion orientations along detector axes (x, y) and the JWST focal plane (V2, V3). The arrows point in the direction of increasing wavelength for R and C grisms in modules A (right) and B (left). Each module has a $2\farcm2 \times 2\farcm2$ imaging field of view.}
\end{figure*}

The NIRCam LW grisms are expected to be used in either of two modes: single-object time-series slitless spectroscopy or wide-field (multi-object) slitless spectroscopy. The layout of the grism dispersions on the NIRCam LW detectors and in the JWST focal plane is shown in Figure~\ref{fig:FP_layout}. A single object can be positioned at the appropriate point to have its entire spectrum (as limited by the chosen filter) fall onto the detector array. The grisms can be operated in series with any of the filters in the NIRCam LW pupil wheel in principle, but several of these would be redundant. Table~\ref{tab:grism_filters} lists all wide and some of the medium-band the filters that will likely be enabled for use with the grisms during JWST observing Cycle 1. We expect those to be most popular, and the remaining medium-band LW filters are also expected to be available for use in wide-field mode in Cycle 1.

Spectra of multiple objects in the imaging field will be recorded, and large sky areas can be mapped with multiple telescope pointings. The DHS elements require that a single object be positioned near a single field point to collect its entire $\lambda = 1 - 2$ $\mu$m spectrum from all 10 DHS elements. Multiple objects in the field can easily produce overlapping DHS spectra, so DHS observations should be restricted to single objects, compact regions, or relatively uncrowded fields (if and when the DHSs are supported for scientific use).

\begin{table}
\centering
\begin{threeparttable}[ht]	
\caption{Filters available for use with LW grisms in Cycle 1} 
\label{tab:grism_filters}    
\begin{tabular}{|l|r|r|r|r|l||} 
\hline
\rule[-1ex]{0pt}{3.5ex}  Filter Name\tnote{a}  & $\lambda_1$ ($\mu$m)\tnote{b} & 
$\lambda_2$ ($\mu$m)\tnote{c} & \# dispersed pixels & \# pixels/2048\tnote{d} & Mode\tnote{e}\\
\hline
F277W  & 2.416 & 3.127 & 711  & 0.35 & TS + WF \\ \hline
F322W2 & 2.430 & 4.013 & 1583 & 0.77 & TS + WF \\ \hline
F356W  & 3.140 & 3.980 & 840  & 0.41 & TS + WF \\ \hline
F444W  & 3.880 & 4.986 & 1106 & 0.54 & TS + WF \\ \hline
F430M  & 4.167 & 4.398 & 231  & 0.11 & WF \\ \hline
F460M  & 4.515 & 4.747 & 232  & 0.11 & WF \\ \hline

\end{tabular}
\begin{tablenotes}
      \small
	  \item [a] All LW M filters will also likely be available in wide-field mode; F430M and F460M are expected to be popular and are included for illustrative purposes
	  \item [b] Half-power wavelength (blue side)
	  \item [c] Half-power wavelength (red side)
	  \item [d] Fraction of the detector that a continuum spectrum occupies in the dispersion direction
	  \item [e] TS = single-object time series and WF = wide field modes
    \end{tablenotes}
\end{threeparttable} 
\end{table}

The NIRCam detectors will be operated in their normal MUTIACCUM
modes\cite{BBK14} during spectroscopic observations. High-precision
time-series grism observations of bright objects (e.g., host stars of
transiting planets) will likely require using a limited detector region
to prevent saturation. Sub-regions consisting of 64, 128, 256, or the
full 2048 detector rows will be supported for single-object time-series
LW grism observations. These can be read out either using all 4 outputs
(stripe mode; spanning all 2048 columns) or else in subarray (or
window) mode using a single output and possibly a smaller number of
columns. Using the combination of stripe mode and 64 detector rows
gives the fastest readout time and provides the best bright limits
(appearing in Table~\ref{tab:grismA}). Subarrays with a smaller
dimension of 64 pixels are also likely to be supported in wide-field
mode, and we expect that these will be used primarily for calibration
purposes.

LW grism spectral observations can be partnered with either simultaneous DHS spectra (if and when DHS use becomes supported) or imaging observations of the same field in the SW channels. SW imaging observations can be conducted in-focus using only filters (in both LW spectroscopic modes), or else out of focus using a weak lens with a filter in series (single-object time-series mode only). Given the flux differences between dispersed spectra and in-focus images, we expect that many single-object point source observers will opt for using a weak lens in series with a filter in the SW channel for simultaneous observations with the LW grisms. 

All NIRCam detectors will be operated with the same readout pattern and
subarray size, so SW observations will use the same size detector
regions as used for the LW grisms. This will limit the number of DHS
spectra recorded unless the full frame readout mode is selected. When
collecting simultaneous SW weak lens (8 waves of defocus at 2.12
$\mu$m) observations, the smallest supported subarray size will be 160
pixels. This is large enough to capture ~98\% of the light from the
source, while providing adequate area around the PSF for background
measurements (190 $\times$ 190 pixel regions collect nearly all light,
requiring use of a larger subarray). As noted earlier, the subarrays
used in the SW and LW channels will have the same dimensions. However,
their locations on the LW and SW detectors can be individually
adjusted. This is of particular importance if simultaneous observations
using the DHSs and LW grisms become a supported mode; it allows the
location of the SW subarrays to be tailored to include the best
centered and brightest DHS spectra while keeping the LW grism spectrum
centered in the LW subarray. When SW data are taken in parallel with
the LW grism data in any mode, particular care must be exercised to
ensure that the single set of exposure parameters provide adequate
dynamic range in both the SW and LW channels.


Short single-object time-series integrations will likely use the RAPID or BRIGHT1 multiaccum modes that save every or every other frame read. This will produce a large amount of data, especially when conducting simultaneous DHS and LW grism observations since they will use three detector arrays (if DHS use is implemented). The on-board solid state data recorder can store 7 -- 10 hours of RAPID and for 3 detector arrays operated in stripe mode (four outputs), and 28 -- 40 hours when operated with a single detector output. JWST is expected to download data every 12 hours, so all but the longest transits of the brightest stars can be observed with both the LW grisms and SW DHSs.

\section{Simulations of Spectroscopic Observations} \label{Simulations}

We have developed software tools to simulate NIRCam spectroscopic observations in order to assess expected performance and to plan using NIRCam in both single-object and multi-object wide-field slitless spectroscopy modes. We describe these tools and present some results in this section.

The first simulator is designed to produce spectral images from multiple objects distributed in 2 spatial dimensions within a given NIRCam field. We have implemented this using the aXeSIM package, first developed for simulating HST WFC3 slitless spectroscopic images\footnote{see http://axe.stsci.edu/axesim/}. Spectra of individual objects can be extracted from simulated images using conventional astronomical data tools. We have produced aXeSIM configuration files that represent the expected signal and noise performance of the NIRCam LW grisms as we expect them to be used on JWST. In addition to producing the wide-field slitless spectroscopy simulations in this section, we have used aXeSIM to produce the NIRCam performance estimates given in Table~\ref{tab:grismA}. These estimates agree with those of SimNRC, the internal NIRCam team exposure time calculator, to within $\sim10$\%. Please contact T. Greene if you would like a package of the NIRCam LWA R grism configuration files for use with aXeSIM (contact information on first page of manuscript). 

We have also developed a second simulator that is used to estimate the signals and noise expected for high precision time series observations of transiting planet spectra. These codes include the same throughputs (e.g., Figure~\ref{fig:tput_and_R}) used in the aXeSIM configuration files, and they also include an expected systematic noise floor in addition to the regular noise terms for detector read noise, dark noise, and observatory background noise components. One-dimensional simulations of exoplanet transmission (in transit) or emission (in secondary eclipse) spectra are produced, and they are described further in a recent scientific publication\cite{GLM16}.

\subsection{Wide-Field Slitless Spectroscopy of Extragalactic Fields}

{\it JWST} is expected to revolutionize the study of high-redshift
galaxies, allowing us to spectroscopically study familiar rest-frame
optical emission lines at high redshifts. With a wavelength coverage of
$2.4 - 5\mu$m, the NIRCam LW grism mode can detect H$\alpha$ at
$z=2.7-6.6$, [O~III] 5007 \AA\ at $z=3.8-9.0$, and [O~II] 3727 \AA\ at
$z=5.4-12.4$. Although NIRSpec will be the preferred choice for most
extragalactic spectroscopic observations because of its superior
sensitivity and wider wavelength coverage ($1-5$ $\mu$m for
$R=1000/2700$; $0.7-5.0$ $\mu$m for $R=100$), the NIRCam LW grism mode
does offer some interesting and potentially powerful capabilities. For
example, the NIRCam LW grism mode will allow: (1) blind searches for
strong line emitters at high redshift, (2) spatially-resolved
spectroscopy for a large number of galaxies simultaneously, and (3)
spectroscopic surveys without pre-imaging data (well suited for
parallel programs).

\begin{figure*}
\begin{center}
\includegraphics[width=1.0\textwidth,angle=0]{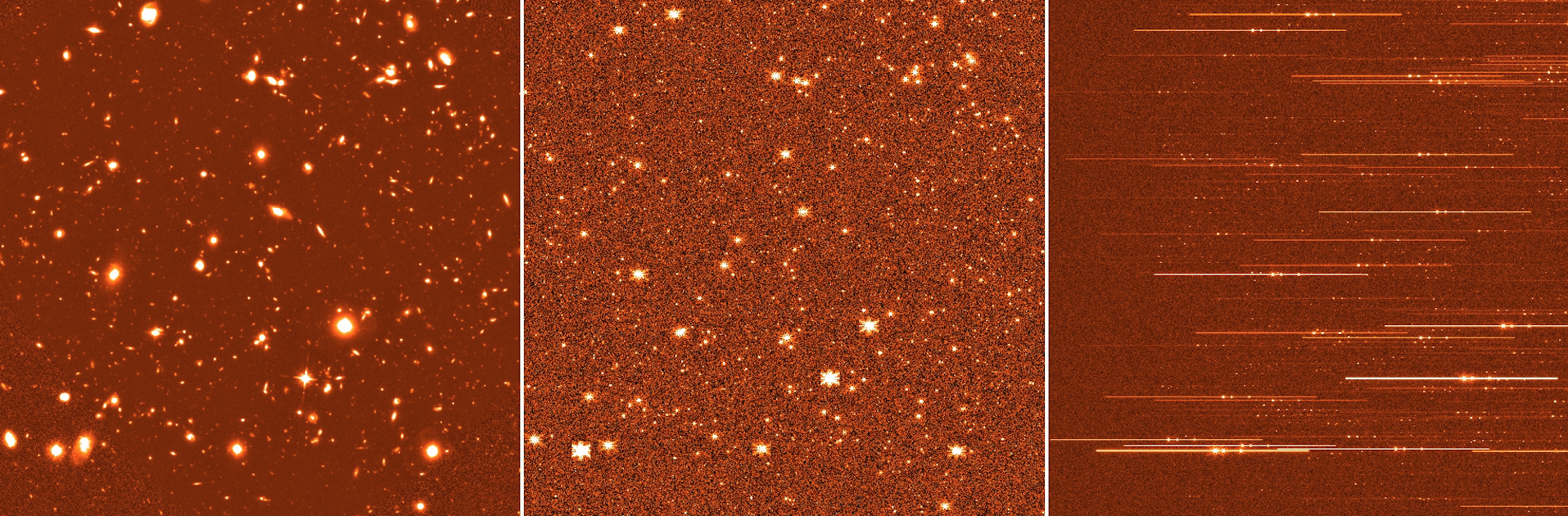}
\end{center}
  \caption{\small Left: {\it HST}/WFC3-IR F160W image of the Hubble
    eXtreme Deep Field\cite{IMO13} (XDF) covering the
    area around the Hubble Ultra-Deep Field (HUDF).  The total
    exposure time is $\sim$65 hours over most of the image; Center:
    2-hour JWST NIRcam image in the F356W filter simulated with the
    source catalog produced from the XDF F160W image on the left.  For
    simplicity, all the sources were assumed to be point-like (i.e.,
    represented by the simulated WebbPSF F356W PSF) and $m_{356, AB} =
    m_{160, AB}$; Right: Simulated 2-hour NIRCam LW grism data in the
    F356W filter and R grism.  The input model spectrum was
    constructed as a combination of a flat $f_{\nu}$ continuum and
    three emission lines, H$\beta$ and [O~III] 4959/5007 \AA\ lines,
    with rest-frame equivalent widths of $\sim$180, 200, and 600 \AA,
    respectively (the line widths were assumed to be unresolved).  All
    the sources were placed at $z=6$ to illustrate the detectability
    of the three emission lines as function of source brightness
    (wavelength increases toward left). \label{xdf}}
\end{figure*}

Figure~\ref{xdf} shows the simulation of a 2-hour NIRCam LW grism
exposure taken over the eXtreme Deep Field\cite{IMO13} (XDF) using the F356W filter.  Here, the redshifts of all the
sources have been set to 6 artificially to illustrate the
detectability of H$\beta$ and [O III] 4959/5007 \AA\ lines as function
of source brightness.  In this image, the continuum is visible down to
$\sim$24 AB mag ($\sim$1 $\mu$Jy) while emission lines are visible
down to $\sim 2\times10^{-18}$ erg cm$^{-2}$ s$^{-1}$.  Although the
confirmation of actual sensitivities has to wait for in-orbit
observations, this simulation clearly demonstrates that NIRCam LW grism
observations of extragalactic fields will produce rich data sets.

Recent studies indicate that the rest-frame optical emission lines of
some $z\sim7$ galaxies are quite large\cite{SBL14}. The combined
rest-frame equivalent width of H$\beta$+[O III] 4959/5007 \AA\ lines
can be well over 1000 \AA, and could reach as high as 2000--3000 \AA\
in some cases (e.g., Smit et al. 2014). This suggests that although the
NIRCam LW grism mode is not as sensitive as NIRSpec, it may be quite
effective in finding high-redshift galaxies through blind search for
strong line emitters (note the high signal-to-noise emission lines in Figure~\ref{xdf}).

\subsection{Wide-field Slitless Spectroscopy of Dark Clouds}

\begin{figure*}
\begin{center}
\includegraphics[width=1.0\textwidth,angle=0]{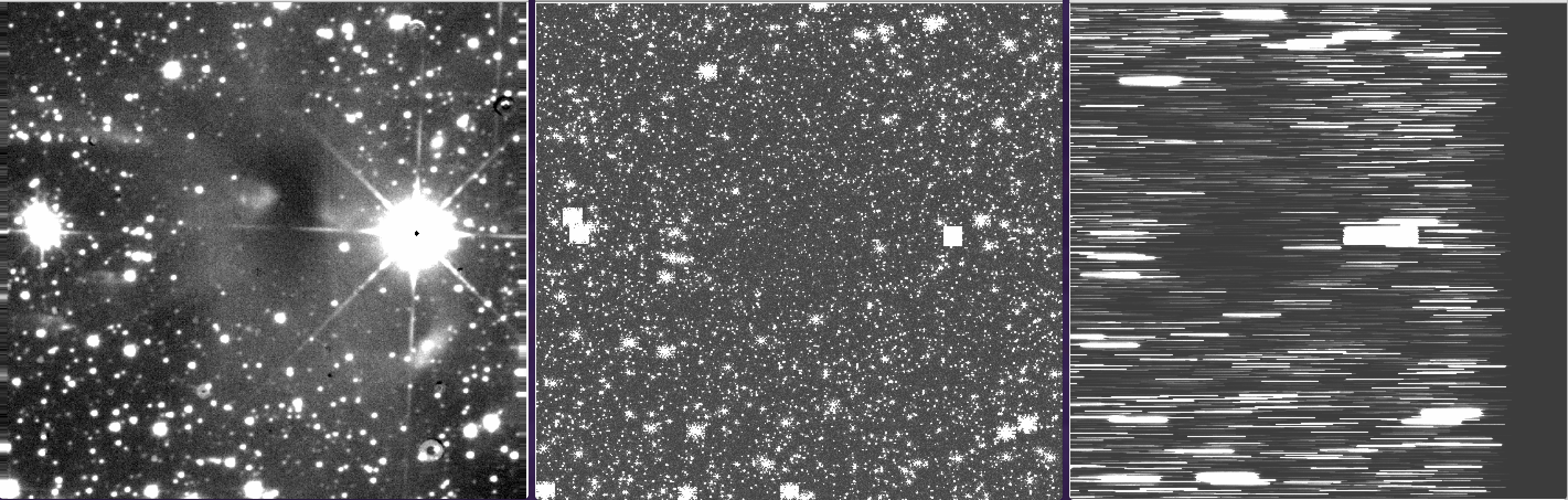}
\end{center}
     \caption{ \label{fig:B335} 
\small 
Left: B335 dark coud $K$-band ($\lambda = 2.2$ $\mu$m) UKIRT image with 3.2 hr integration time. Center: Simulated 1 second JWST NIRCam image in the F430M filter, using simulated WebbPSF PSFs (see http://www.stsci.edu/wfirst/software/webbpsf) for the stars seen in the UKIRT data and also added fainter stars. Right: Simulated 30 second NIRCam F430M R grism image, showing the spectra of the individual stars seen in the images. Background stars in high extinction regions show little overlap in their relatively short spectra, and most overlaps can be resolved with additional observations taken with the C grism. This spectral region will show 4.3 $\mu$m CO$_2$ absorptions from ice mantles forming on dust grains in the cloud.
}
\end{figure*}

Sun-like stars form in nearby dark clouds within our galaxy; the gas and dust within these regions is pulled together by gravity until the central regions form stars. These gravitationally-confined dark globules are very dense and emit no visible light, but they can be studied using the infrared light of background stars. The amount of dust (silicate and carbon) and gas material in these clumps can be measured by observing the brightness of these background stars at several different infrared wavelengths using NIRCam filter imaging. Likewise the amount of H$_2$O, CO, and CO$_2$ ice on the cloud's dust grains can be measured with NIRCam LW grism spectra, as simulated in Figure~\ref{fig:B335} for the B335 cloud. Background stars in high extinction regions show little overlap in their relatively short spectra through F430M (CO$_2$) or F460M (CO) filters, and most overlaps can be resolved with additional observations taken with the grism having dispersion in the perpendicular direction. Observations like these will show how the density of star-forming clouds changes with their radius and how volatile materials freeze out onto dust grains.

\subsection{Spectra of Transiting Planets} \label{transits}

\begin{figure*}
\begin{center}
\includegraphics[width=1.0\textwidth,angle=0]{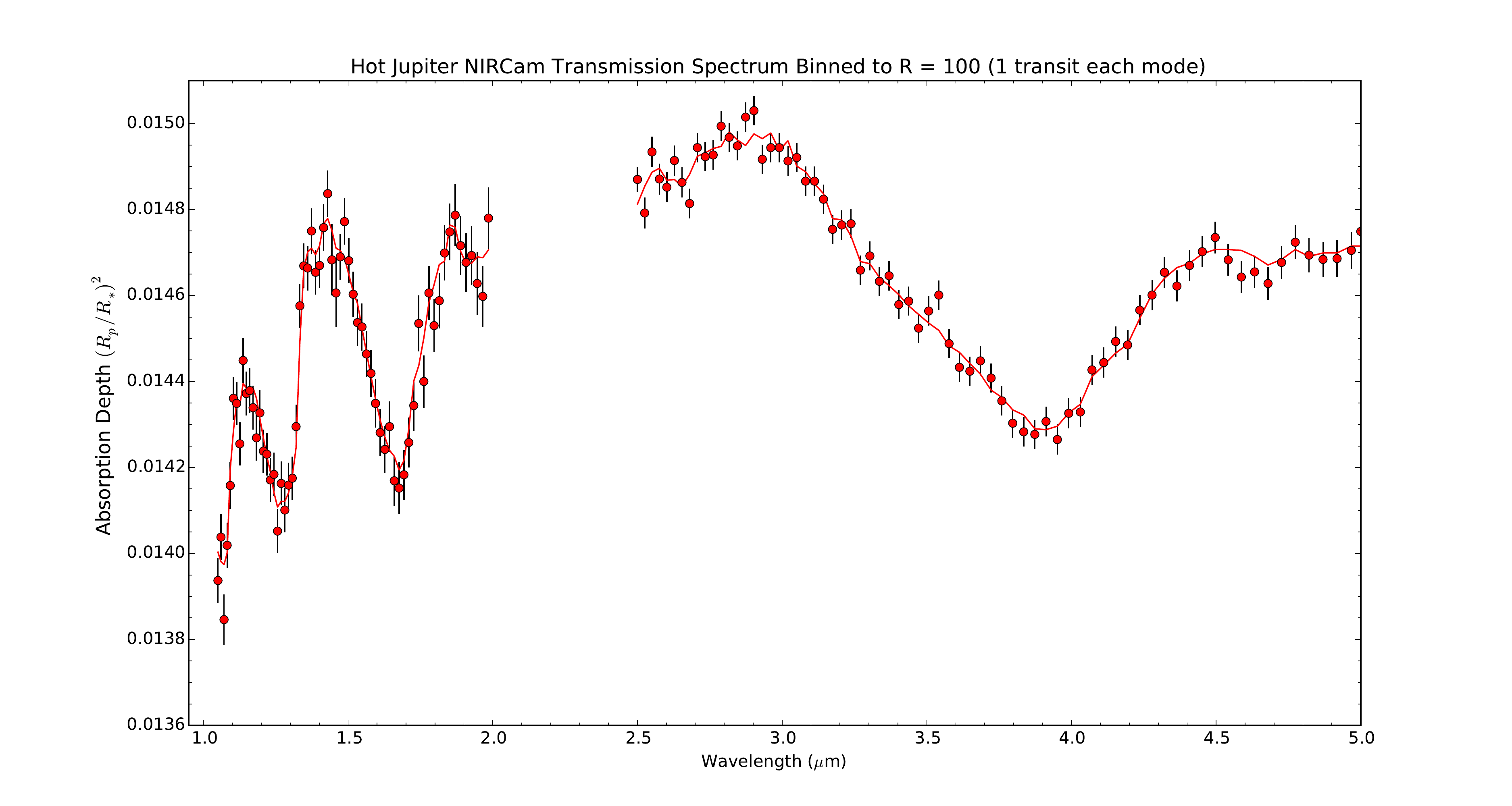}
\end{center}
     \caption{ \label{fig:HD209} 
\small  Simulated NIRCam DHS plus LW grism transmission observation of a model hot Jupiter planet with a clear solar composition atmosphere and the system parameters of HD 209458b. The transmission model is shown as the solid red curve, and simulated data are shown as points with error bars denoting uncertainties. A single transit plus equal time on the star alone was simulated at each wavelength, and 2 DHS spectra are co-added, appropriate for a 256 row detector subarray. Covering this spectral range would require observations of at least 2 transits (see Section \ref{transits}). We note that DHS use is not yet supported for scientific observations.}
\end{figure*}

We have used our 1-D spectral time-series simulator to model the
expected JWST spectra of several transiting planets and estimate their
scientific information content\cite{GLM16}. We have recently extended
this code to include simulations of SW DHS data, and we show an example
of a combined SW DHS and LW grism spectrum in Figure~\ref{fig:HD209}.
The model transmission spectrum of a hot Jupiter planet is shown as a
solid red curve, and the simulated data have been binned to
spectroscopic resolution R = 100 and are shown as red data points with
error bars indicating their uncertainties. These observations would
have to be made during at least two separate transits: one transit with
a grism plus F322W2 filter, one with a grism and the F444W filter, and
SW DHS observations made simultaneously with at least one of the LW
grism exposures. Strong absorptions (large absorption depth values) of
H$_2$O are clearly seen at 1.2, 1.4, 1.9, and 3.0 $\mu$m, with CO at
4.6 $\mu$m. These observations should constrain the H$_2$O and CO
mixing fractions of similar transiting planets to factors of 2 -- 3,
much better than HST or other current observations\cite{GLM16}.

\section{Observation Planning} \label{Planning}

We intend for the information provided above to be sufficient for
assessing the basic feasibility of most NIRCam slitless spectroscopic
observations. Planning tools for detailed assessment of sensitivity in
specific instrument configurations (as explained in previous sections)
and for designing observation sequences will be available through the
Space Telescope Science Institute (STScI), the JWST science operations center. Simultaneous SW imaging and LW grism observations will be supported for Cycle 1 observations, but the schedule for implementing
simultaneous SW DHS plus LW grism observations is not currently known.

A prototype exposure time calculator (ETC) is currently available for
estimating sensitivity and needed exposure times, but it does not
support the LW grism mode or imaging through the SW weak lens. A new,
scene-based ETC (Pandeia) will support LW grism and all other major
JWST observation modes when it is released in January 2017. As
described in Section~\ref{Simulations}, we can also provide
NIRCam-specific configuration files for the aXeSIM package that can be
used to simulate NIRCam LW grism observations and estimate their
signal-to-noise. Once the needed integration time has been calculated
(e.g., from Pandeia, aXeSIM, or Table~\ref{tab:grismA}), observation
details can be planned with the Astronomer's Proposal Tool
(APT\footnote{see http://www.stsci.edu/hst/proposing/apt}) for JWST.
The current version of APT includes a template for the grism
single-object time-series observation mode, including simultaneous
imaging in the SW channel. The template for planning wide-field
slitless spectroscopy observations is being designed now, and will be
available by early 2017. These templates allow selection of targets,
detector readout (and subarray) parameters, filters, grisms, and (for
the wide-field mode) mosaics and dithers. APT provides estimates of
observing windows for targets (consistent with the nominal launch date,
orbit, and field of regard restrictions), as well as observing
overheads.

\acknowledgments 
 
We thank D. Coe for preparing the figure showing the layout of grism spectra in Figure~\ref{fig:grism} and S. Lilly for discussions and assistance with determining the LW grism dispersion orientations. We also thank M. Line for the exoplanet transmission model and N. Lewis for operations information.
TPG and coauthors acknowledge funding support by the NASA JWST project for NIRCam, NASA WBS 411672.05.05.02.02.

\bibliography{nircam_spec} 
\bibliographystyle{spiebib} 

\end{document}